\def\sqr#1#2{{\vcenter{\vbox{\hrule height.#2pt\hbox{\vrule
width.#2pt height#1pt \kern#1pt\vrule width.#2pt}\hrule height.#2pt}}}}
\def\square{\mathchoice\sqr54\sqr54\sqr{2.1}3\sqr{1.5}3} 

\documentclass[12pt,a4paper]{article} 

\font\cst=cmr10 scaled \magstep4
\font\csc=cmr10 scaled \magstep2

\begin{document}

\vglue 1cm
\centerline{\cst On the quasi-linearity of the}
\vskip 0.5cm
\centerline{\cst Einstein ``Gauss-Bonnet"  gravity field equations}
\vskip 1.2cm

\centerline{\bf Nathalie Deruelle}
\vskip 0.2 cm
\centerline{\it  Institut d'Astrophysique de Paris,}
\centerline{\it GReCO, FRE 2435 du CNRS,}
\centerline{\it 98 bis Boulevard  Arago, 75014 Paris, France}

\centerline{and}

\centerline{\it Yukawa Institute for Theoretical Physics}
\centerline{\it Kyoto University, Kyoto 606-8502, Japan}

\vskip 0.5 cm
\centerline{\bf John Madore}
\vskip 0.2cm
\centerline{\it Laboratoire de Physique Th\'eorique,}
\centerline{\it  UMR 8627 du CNRS, Universit\'e de Paris XI}
\centerline{\it 91405 Orsay, France}

\medskip
\vskip 0.5cm
\centerline{April 2003}

\vskip 1cm

\centerline{\it Dedicated to Yvonne Choquet-Bruhat.}
\vskip 2cm

\noindent
{\bf Abstract}
\bigskip

\noindent
We review some properties of the  Einstein ``Gauss-Bonnet" equations for gravity---also called the Einstein-Lanczos equations in five and six dimensions, and the
Lovelock or Euler equations in higher dimensions. We illustrate, by means of simple Kaluza-Klein and brane cosmological models, some consequences of the quasi-linearity of
these equations on the Cauchy problem  (a point first studied by  Yvonne Choquet-Bruhat), as well as on ``junction conditions".
\vfill\eject

\noindent
{\csc I. Introduction}

\medskip
Vermeil~\cite{Ver17} as early as 1917, then  Weyl~\cite{Wey21} and Cartan~\cite{Car22}, showed that the Einstein tensor was, in any dimension, the only symmetric and
conserved  (that is divergence free) tensor depending only on the metric, its first and second derivatives, the dependence in the latter being linear (see below and
e.g.~\cite{ChoDew82} for a definition of (quasi)-linearity). 

Lanczos~\cite{Lan32} in 1932 found a generalization of Hilbert's Lagrangian  which is  quadratic in the Riemann tensor and yields, via a Euler variation with
respect to the metric, a tensor (the Lanczos tensor)  which is, like Einstein's, symmetric, conserved and second order in the metric but, in contrast to Einstein's, only
quasi-linear in these second derivatives. Another important property is that, just like Hilbert's Lagrangian is  a pure  divergence in two dimensions and Einstein's tensor
identically zero in one and two dimensions, we have that in four dimensions the Lanczos Lagrangian is a pure divergence (a  Euler density, or ``Gauss-Bonnet",
term~\cite{ChoDew82}~\cite{KobNom69}) and that the Lanczos tensor is identically zero in four or less dimensions (a property already known to Bach~\cite{Bac21}). 

Lovelock~\cite{Lov71} generalized these results in 1971 and obtained, for any dimension, a formal expression for the most general, symmetric and conserved
tensor which is quasi-linear in the second derivatives of the metric and does not contain any higher derivatives. He also found the Lagrangian from which that tensor derives by
a Euler variation with respect to the metric~: in dimension $D$ it is, in fact, a linear combination of the  $[D/2]$ dimensionally continued Euler densities~;  hence, in dimensions 5
and 6, the explicit form of the Lovelock Lagrangian reduces to a  linear combination of Hilbert's and Lanczos' Lagrangians~; in dimensions 7 and 8 its explicit expression as well
as that of the  derived tensor (which are  cubic  in the curvature) were given by M\"uller-Hoissen~\cite{Mul86}~; an expression for the developped form of the quartic Lagrangian,
which is the highest to come into play in 9 and 10 dimensions can be found in ref~\cite{Whe86}. Given its properties, the Lovelock Lagrangian is therefore the most natural
generalization of Hilbert's to describe pure gravity in higher dimensional spacetimes.

The idea now that space-time may have more than four dimensions has been recurrent in unified field theories since the original proposal by Kaluza~\cite{Kal21} and
Klein~\cite{Kle26}, (see e.g.~\cite{Lic55} for an account of the ``classical period", and~\cite{AppChoFre87} for a more recent perspective)---and since the renewal
of string theories  (see e.g.~\cite{GreSchWit87}), it is considered as almost a fact. As for modifying, in four dimensions,  the Hilbert Lagrangian by the inclusion of terms that
are non-linear in the curvature, it is an idea that goes back to Weyl~\cite{Wey21} and Eddington~\cite{Edd24}~; of course, from the theorems mentionned above, the
Euler-derived tensors (symmetric and conserved)  contain (in order not to be identically zero in four dimensions) terms in derivatives of the metric up to the fourth.  In the
seventies and early eighties such quadratic Lagrangians were exploited in view of renormalizing the quantized theory of linearized general relativity (see e.g.~\cite{Bou84} for
a review of that period) as well as to renormalize the stress-energy tensor of quantized matter fields in classical, curved, backgrounds, see~\cite{BirDav82} for a review. They
however made their most forceful entrance when it was shown that they should arise from string theories, see~\cite{GreSchWit87}.

The Lovelock tensor differs however from the tensor derived from a generic non-linear correction to the Hilbert Lagrangian in that it contains derivatives of the metric
of order no higher than the second. The main consequence of this property is to avoid the appearance, by increasing the order of differentiation of
the field equations, of new classes of dynamical solutions which do not approach the unperturbed, Einsteinian, solutions when the non-linear terms in the Riemann tensor tend
to zero. Hence the Lovelock tensor is of little use in achieving goals which ultimately relie on the possibility of singular perturbations such as the renormalization of the
graviton propagator~\cite{Ste77} or the type of inflation first discussed by Starobinski~\cite{Sta80}. However the same property guarantees that the quantization of the
linearized Lovelock theory is free of ghosts and it was argued that, for this reason, the Lovelock Lagrangian should appear in the low-energy limit of string theories
(see~\cite{Zwi85}, and, e.g.~\cite{GroSlo87}).

For all these reasons, gravity theories based on the Lovelock tensor were extensively investigated in the late 80's, starting with the work of Madore~\cite{Mad85a} and
M\"uller-Hoissen~\cite{Mul85} (see e.g.~\cite{Der87a} for a review). In particular, the consequences of the quasi-linearity of these equations on the Cauchy problem and
structure of characteristics were studied by  Choquet-Bruhat in~\cite{Cho88a} (see also~\cite{Ara88}). The closely connected problem of the wave propagation was examined
in~\cite{GibRub86}, as well as the resulting difficulties in setting up a Hamiltonian formalism~\cite{TeiZan87}. As for the linear stability of various candidate ground states,
see e.g.~\cite{BouDes85}. Finally, the general setting for establishing conservation laws was given in~\cite{DubMad87}.

Cosmological models also  became a prime focus of interest as the early universe appeared to be a privileged arena where unified theories could be probed (see
e.g.~\cite{Mad85a}~\cite{Mul85}~\cite{DerMad86a}~\cite{Ish86} for a taste of the results then obtained, and~\cite{BaiLovWon85} when gravity
is coupled to various other gauge matter fields). Interest in the subject then faded away, partly perhaps  because  priority was given to studying the would-be observable
properties of the string-theory-predicted dilaton---which are better described by four dimensional scalar-tensor theories of gravity (see e.g.~\cite{DamEsp95}).

We are however at present witnessing a comeback of  pure gravity theories based on the Lovelock (or, rather, Lanczos) tensor which is motivated by the invention of  ``brane
scenarios"~\cite{RanSun99a} and, in particular, of cosmological models in which the observable universe is described as a four dimensional singular surface, or ``brane", of a
five dimensional space-time, or ``bulk" (see e.g.~\cite{BraBru03} and references therein).  For an introduction to the current literature on Gauss-Bonnet gravity in braneworlds,
see~[38-55].

In this contribution we shall review some properties of the Lovelock field equations---more specifically the five dimensional Einstein-Lanczos equations (very often called
Einstein  Gauss-Bonnet) and focus on some consequences of their quasi-linearity, that we shall illustrate by ``paradigmatic" Kaluza-Klein or brane cosmological models.  We shall
also review a closely related question which has been debated recently, that is the generalized ``junction conditions"~\cite{Isr66} in Lanczos gravity.

\bigskip

\noindent
{\csc II. The Lanczos and Lovelock Lagrangians and tensors}

\bigskip

\centerline{\bf a. Quadratic and Lanczos-Gauss-Bonnet Lagrangians}

Consider, in some coordinate system $x^A$, the quadratic Lagrangian density
$$ \sqrt{-g}\,L_q\equiv\sqrt{-g}\left(s^2+\beta\, r^{LM}r_{LM}+\gamma\, R^{LMNP}R_{LMNP}\right)\eqno({\rm II}.1)$$
where $g$ is the determinant of the  metric coefficients $g_{AB}$, and where $R^A_{\ BCD}\equiv\partial_C\Gamma^A_{BD}-...$ are the components of the  Riemann tensor, 
$\Gamma^A_{BD}$ being the Christoffel symbols, all  indices being raised with the inverse metric $g^{AB}$~;
$r_{BD}\equiv R^A_{\ BAD}$ are the Ricci tensor components, $s\equiv g^{BD}r_{BD}$ is the scalar curvature, and $\beta$ and $\gamma$ are some constants. 
Varying this density with respect to  $g_{AB}$ is a standard calculation which gives
$$\delta (\sqrt{-g}\,L_q)=\sqrt{-g}\, H_{AB}\,\delta g^{AB}+\sqrt{-g}D_CV^C\eqno({\rm II}.2)$$
where
$$H_{AB}\equiv 2\gamma R_{ALNP}R_B^{\ LNP}-2(2\gamma+\beta)r^C_DR^D_{\ BAC}-4\gamma r_A^Cr_{CB}+2
sr_{AB}-{1\over2}g_{AB} L_q+$$
$$+ (\beta+4\gamma)\square r_{AB}+{1\over2}(4 +\beta)g_{AB}\square s-(2 +\beta+2\gamma)D_{AB}s\eqno({\rm II}.3)$$
and
$$V^C\equiv\left\{(4\gamma+\beta)D^Cr^{BD}+{1\over2}(4 +\beta)g^{BD}D^Cs -2D^B\left[(\beta+2\gamma)r^{CD}+g^{CD}s\right] \right\}\delta g_{BD}+\eqno({\rm II}.4)$$
$$+[2s\,(g^{AB}g^{CD}-g^{AC}g^{BD})+\beta(2r^{AB}g^{CD}-r^{BD}g^{AC}-r^{AC}g^{BD})+4\gamma R^{BACD}]D_A\delta g_{BD}$$
$D$ being the covariant derivative associated with  $g_{AB}$ and where $\square\equiv g^{CD}D_{CD}$. Note that
$V^C$ is defined up to the addition of the divergence-less vector $2c\, U^C$ with $c$ an arbitrary constant and
$$U^C\equiv{1\over2}D_A[(r^{AB}g^{CD}-r^{BC}g^{AD}+r^{AD}g^{BC}-r^{CD}g^{AB})\delta g_{BD}]=\eqno({\rm II}.5)$$
$$=D_A[(r^{AB}g^{CD}-r^{DC}g^{AB})\delta g_{BD}]=-(D^B\sigma^{DC})\,\delta g_{BD}+(r^{AB}g^{CD}-r^{BC}g^{AD})D_A\delta g_{BD}$$
$\sigma^{DC}\equiv r^{DC}-{1\over2}sg^{DC}$ being the contravariant components of the Einstein tensor.\footnote{Recall that the variation of the Hilbert Lagrangian
density $\sqrt{-g}\,s$ with respect to the metric coefficients is $\delta (\sqrt{-g}\, s)=\sqrt{-g}\,\sigma_{AB}\,\delta g^{AB}+\sqrt{-g}\,D_CV^C_{(1)}$ where
$\sigma_{AB}\equiv r_{AB}-{1\over2}sg_{AB}$ are the covariant components of the Einstein tensor, and where $V^C_{(1)}=(g^{AB}g^{CD}-g^{AC}g^{BD})D_A\delta g_{BD}$ is
uniquely defined.}

\medskip
The curvature tensor components being second order in the derivatives of the metric coefficients, the Lanczos Lagrangian, $L_{(2)}$, is defined by the non trivial combination
$\beta=-4$, $\gamma=1$, 
such that the corresponding Lanczos tensor (of mixed components denoted $\sigma^A_{(2)B}$), is only second order in the derivatives of the metric
coefficients~\cite{Lan32}~: 
$$L_{(2)}\equiv s^2-4r^{LM}r_{LM}+R^{LMNP}R_{LMNP}\eqno({\rm II}.6)$$
$$\sigma^A_{(2)B}\equiv 2\left[R^{ALMN}R_{BLMN}-2r^{LM}R^A_{\ LBM}-2 r^{AL}r_{BL}+
sr^A_B\right]-{1\over2}\delta^A_B L_{(2)}\,.\eqno({\rm II}.7)$$
 Since $H_{AB}$ and $\sigma^A_{(2)B}$ derive from  a Lagrangian they are conserved~:
$$D_AH^A_B= 0\qquad,\qquad D_A\sigma^A_{(2)B}=0\eqno({\rm II}.8)$$
for any metric $g_{AB}$ (as an explicit calculation, using the Bianchi identities, confirms). As for the boundary term (4) it can be written, upon a proper choice of the divergenless vector (5),
as~\cite{Dav02}
$$V^C_{(2)}=-4P^{ABCD}D_A\delta g_{BD}\eqno({\rm II}.9)$$
with 
$$P^{ABCD}\equiv R^{ABCD} +(r^{AD}g^{BC}-r^{BD}g^{AC}+r^{BC}g^{AD}-r^{AC}g^{BD})-{1\over2}s\,(g^{AB}g^{CD}-g^{AC}g^{BD})\,.$$

A pedestrian way  to see (and define) the quasi-linearity of $\sigma^A_{(2)B}$ (which has to be taken anyhow when one goes to practical calculations) is to introduce a (local) 
Gaussian normal 
coordinate system ($x^A=\{w,x^\mu\}$) such that the line element reads (we use the mostly $+$ signature)
$$ds^2=\epsilon \, dw^2+\gamma_{\mu\nu}\,dx^\mu dx^\nu\,.\eqno({\rm II}.10)$$
 The surface $w=Const.$ is time- or space-like if $\epsilon=+1$ or $-1$ and, in terms of the components of its extrinsic curvature 
$$K_{\mu\nu}\equiv -{1\over2}{\partial\gamma_{\mu\nu}\over\partial w}\eqno({\rm II}.11)$$
the components of the Riemann tensor decompose into the gaussian normal version of the Gauss-Codazzi-Mainardi equations, that is  :
$$R_{w\mu w\nu}={\partial K_{\mu\nu}\over\partial w}+K_{\rho\nu}K^\rho_\mu\quad,\quad R_{w\mu\nu\rho}=\nabla_\nu K_{\mu\rho}-\nabla_\rho K_{\mu\nu}$$
$$R_{\lambda\mu\nu\rho}=\bar R_{\lambda\mu\nu\rho}+\epsilon[K_{\mu\nu}K_{\lambda\rho}-K_{\mu\rho}K_{\lambda\nu}]\eqno({\rm II}.12)$$
where $\nabla_\rho$ and $\bar R^\mu_{\ \nu\rho\sigma}$ are the covariant derivative and the Riemann tensor associated with the metric $\gamma_{\mu\nu}$, all Greek indices
being raised with the inverse metric $\gamma^{\mu\nu}$.  It is then an easy exercise to see that, first, $\sigma^A_{(2)B}$ does not contain terms  in $(\partial
K^\mu_\nu/\partial w)^2$ (as one would naively expect since 
$\sigma^A_{(2)B}$ contains terms in $(R_{LMNP})^2$), second,  there are no terms linear in
$(\partial K^\mu_\nu/\partial w)$ in $\sigma^w_{(2)w}$ and $\sigma^w_{(2)\mu}$, and, third,
$$\sigma^{\mu\nu}_{(2)}=N^{\mu\alpha\nu\beta}\,{\partial K_{\alpha\beta}\over\partial w}+...$$
with
$$ N^{\mu\alpha\nu\beta}\equiv 2(K.
K-K^2)(\gamma^{\mu\alpha}\gamma^{\nu\beta}-\gamma^{\mu\nu}\gamma^{\alpha\beta})+4(\gamma^{\alpha\beta}K^{\mu\rho}K^\nu_\rho +\gamma^{\mu\nu}K^\alpha_\rho
K^{\rho\beta})-\eqno({\rm II}.13)$$
$$-4K(\gamma^{\mu\nu}K^{\alpha\beta}+\gamma^{\alpha\beta}K^{\mu\nu})+
8KK^{\mu\alpha}\gamma^{\nu\beta}-8K^{\mu\rho}K_\rho^\alpha\gamma^{\nu\beta}
+4(K^{\mu\nu}K^{\alpha\beta}-K^{\alpha\mu}K^{\nu\beta})-4\epsilon\bar  P^{\mu\alpha\nu\beta}$$
where $\bar P^{\mu\nu\rho\sigma}$ is the restriction onto the surface $w=Const.$ of $P^{ABCD}$ defined in (9);
where $K. K\equiv K^\alpha_\beta K^\beta_\alpha$ and where the $(...)$ stand for terms of zeroth order in $(\partial K^\mu_\nu/\partial w)$. Since $N^{\mu\alpha\nu\beta}$ contains not  
only the metric but also its first derivatives with respect to $w$,  the Lanczos tensor is, contrarily to Einstein's, only
{\it quasi}-linear~\cite{ChoDew82} in the second derivatives of the metric coefficients.\footnote{Recall that in Gaussian normal
coordinates the components $\sigma^w_w$ and $\sigma^w_\mu$ of the Einstein tensor do not depend on the
$w-$derivatives of the extrinsic curvature and that $\sigma^{\mu\nu}=(\gamma^{\mu\alpha}\gamma^{\nu\beta}-\gamma^{\mu\nu}\gamma^{\alpha\beta})\,{\partial
K_{\alpha\beta}\over\partial w}+...={\partial K^{\mu\nu}\over\partial w}-\gamma^{\mu\nu}{\partial K\over\partial w}+...$ is {\it linear} in them, in the sense that the projector 
$(\gamma^{\mu\alpha}\gamma^{\nu\beta}-\gamma^{\mu\nu}\gamma^{\alpha\beta})$ contains the metric coefficients only~\cite{ChoDew82}.} We note for further reference
that, using the Leibniz rule,  the above expression can be simplified into~\cite{DerDoz00}~\cite{Dav02}
$$\sigma^\mu_{(2)\nu}=4{\partial\over\partial w}\left\{{3\over2}J^\mu_\nu-{1\over2}\delta^\mu_\nu\, J -\epsilon\bar P^\mu_{\
\rho\nu\sigma}K^{\rho\sigma}\right\}+...\eqno({\rm II}.14)$$
with
$$ J^{\mu\nu}\equiv -{2\over3}K^{\mu\rho}K_{\rho\sigma}K^{\sigma\nu}+{2\over3}K\,K^{\mu\rho}K_\rho^\nu+{1\over3}K^{\mu\nu}(K.K-K^2)\,.$$

\medskip

\centerline{\bf b. The Lovelock generalisation}

\medskip

In order now to generalize these results and gain further insight into them, one notices that the  Hilbert and Lanczos Lagrangians can
be rewritten as
$$L_{(1)}\equiv s={1\over2}\delta^{I_1I_2}_{J_1J_2}\ R^{J_1J_2}_{\ \ \ \ I_1I_2}$$
$$L_{(2)}={1\over4}\delta^{I_1I_2I_3I_4}_{J_1J_2J_3J_4}\ R^{J_1J_2}_{\ \ \ \ 
I_1I_2}\ R^{J_3J_4}_{\ \ \ \ I_3I_4}\eqno({\rm II}.15)$$
where $\delta^{I_1...I_{2p}}_{J_1...J_{2p}}$ is the Kronecker symbol of order $2p$~: $\delta^{I_1...I_{2p}}_{J_1...J_{2p}}={\rm det}\,\delta^I_J$, $I$ and $J$ standing for
$I_1,...I_{2p}$ and $J_1,...J_{2p}$  (thus, for example~: $\delta^{I_1I_2}_{J_1J_2}=\delta^{I_1}_{J_1}\delta^{I_2}_{J_2}-\delta^{I_1}_{J_2}\delta^{I_2}_{J_1}$). As for the
Einstein and Lanczos tensor, they can also be rewritten in a similar fashion
$$\sigma^A_{(1)B}\equiv r^A_B-{1\over2}\delta^A_B\,s\,=-{1\over4}\delta^{A\,I_1I_2}_{B\,J_1J_2}\ R^{J_1J_2}_{\ \ \ \ I_1I_2}$$
$$\sigma^A_{(2)B}=-{1\over8}\delta^{A\,I_1I_2I_3I_4}_{B\,J_1J_2J_3J_4}\ R^{J_1J_2}_{\ \ \ \
I_1I_2}\ R^{J_3J_4}_{\ \ \ \ I_3I_4}\,.\eqno({\rm II}.16)$$

Lovelock~\cite{Lov71} hence generalized the above definitions and showed that the Euler variation of the Lagrangian density $\sqrt{-g}\, L_{(p)}$ with
$$ L_{(p)}={1\over2^p}\,\delta^{I_1...I_{2p}}_{J_1...J_{2p}}\ R^{J_1J_2}_{\ \ \ \ I_1I_2}\ ...\ R^{J_{2p-1}J_{2p}}_{\ \ \ \ \ \ \ \ \ I_{2p-1}I_{2p}}\eqno({\rm II}.17)$$
was, up to a pure divergence term~:
$$\sigma^A_{(p)B}=-{1\over2^{p+1}}\,\delta^{A\,I_1...I_{2p}}_{B\,J_1...J_{2p}}\ R^{J_1J_2}_{\ \ \ \ I_1I_2}\ ...\ R^{J_{2p-1}J_{2p}}_{\ \ \ \ \ \ \ \ \ I_{2p-1}I_{2p}}\,.\eqno({\rm
II}.18)$$ 
The developped expressions for $L_{(3)}$ and $\sigma^A_{(3)B}$ were obtained in~\cite{Mul86} and an explicit form for $L_{(4)}$ can be found in~\cite{Whe86}.

The Kronecker symbol $\delta^{AI_1...I_{2p}}_{BJ_1...J_{2p}}$ is zero if two or more of, say, its upper $I$ indices are the same~;  it is therefore identically zero in space-times
of dimension $D\leq 2p$~; hence we recover that the Einstein tensor is identically zero in one or two dimensions, and see that the Lanczos tensor (7)  vanishes identically in
four and less dimensions (a property first discovered by Bach~\cite{Bac21}). The Lovelock Lagrangian and tensor are then defined, in a $D$-dimensional space-time, as
$$L_{[D]}=\sum_{0\leq p\leq D/2}\alpha_p\lambda^{2(p-1)} L_{(p)}\qquad,\qquad\sigma^A_{[D]B}=\sum_{0\leq p< D/2}\alpha_p\lambda^{2(p-1)}\sigma^A_{(p)B}\eqno({\rm
II}.19)$$ where we have set $L_{(0)}=1$ and $\sigma^A_{(0)B}=-{1\over2}\delta^A_B$, where $\lambda$ is a length scale (e.g. the Planck scale) and $\alpha_p$ are dimensionless
parameters, which, for lack of a metatheory or observations, have for the time being to be left unspecified.

\bigskip

\centerline{\bf c. The Lovelock Lagrangian and tensor in Cartan's formalism}

The Lagrangians $L_{(p)}$ and tensors $\sigma^A_{(p)B}$ can also be expressed in terms of Cartan moving-frame formalism otherwise known, in dimension four, as the Vierbein
or tetrad formalism, see e.g.~\cite{ChoDew82}. The notation is adapted from~\cite{DubMad87}. As usual we first define an orthonormal frame, that is, a set of $D$ one-forms
$\theta^A$ which are linear combinations of the differential forms ${\rm d} x^A$ naturally associated with the coordinates $x^A$, and  chosen so that the metric can be expressed in the
 form
 $$g \equiv g_{AB}\, {\rm d}x^A\otimes  {\rm d}x^B=\eta_{AB}\, \theta^A\otimes \theta^B \eqno({\rm II}.20)$$
where $\eta_{AB}=(-1, +1,...,+1)$ and where the product can be thought of as the symmetric tensor product.

In place of the Christoffel symbols, we then introduce a connection, that is $D^2$ one-forms $\omega^A{}_B$ which define the covariant derivative by
$$ D \theta^A = - \omega^A{}_B \otimes \theta^B\eqno({\rm II}.21)$$
which is extended to arbitrary forms by the Leibniz rule. The connection is torsion-free and metric, conditions which can
be imposed respectively by the equations
$${\rm d}\theta^A + \omega^A{}_B\,\wedge\, \theta^B = 0, \qquad \eta_{AC}\,\omega^C_{\ B}+\eta_{BC}\,\omega^C_{\ A}=0\,.\eqno({\rm II}.22)$$
The product here is the exterior (anti-symmetrised tensor) product of two forms and ${\rm d}$ denotes the exterior derivative of a form. 

 The curvature 2-form (Cartan's second structural equation)
$$\Omega^A{}_B = {\rm d} \omega^A{}_B + \omega^A{}_C\,\wedge\, \omega^C{}_B\eqno({\rm II}.23)$$
defines the components of the Riemann tensor by the expansion
$$\Omega^A{}_B = {\scriptstyle{1\over2}} R^A{}_{BCD}\ \theta^C \theta^D\eqno({\rm II}.24)$$ 
where the exterior product symbol $\wedge$ is from now on omitted.

An important operation is the duality map. We define for each integer $p$ the $(D-p)$-form 
$$\theta^*_{I_1...I_p}={1\over(D-p)!}\,\epsilon_{I_1...I_D}\theta^{I_{p+1}}... \ \,\theta^{I_D}\eqno({\rm II}.25)$$
where $\epsilon_{I_1...I_D}$ is the completely antisymmetric tensor normalized to $\epsilon_{0...D-1}=1$.
The Hilbert action is then an integral over space-time
of the $D$-form
$${\cal L}_{(1)} = \Omega^{AB} \theta^*_{AB}\,.\eqno({\rm II}.26)$$
The product here is as always the product in the algebra of forms. As discussed above, for each $p$, the Lovelock $D$-forms
$${\cal L}_{(p)} = (\Omega^p)^{I_1 \cdots I_{2p}} \theta^*_{I_1 \cdots I_{2p}}\eqno({\rm II}.27)$$
have similar properties. By $\Omega^p$ we mean $p$ (exterior) products  of the curvature 2-form.

The Lovelock  (D-1)-forms $E_{(p)A}$ are derived by variation of ${\cal L}_{(p)}$ with respect to the frame. Using the Bianchi identities (obtained by exterior differentiation of
 (23)) one sees that this amounts to
simply removing one of the frame factors. Hence~:
$$E_{(p)A} = (\Omega^p)^{I_1 \cdots I_{2p}} \theta^*_{AI_1 \cdots I_{2p}}\,.\eqno({\rm II}.28)$$
The components of the Einstein tensor $\sigma_{AB}$ are then given by
$$E_{(1)A} =-{1\over2} \Omega^{BC} \theta^*_{ABC} = \sigma_{AB} \theta^{*B}   \eqno({\rm II}.29)$$
(where $\theta^{*A}=\eta^{AB}\theta^*_B$). 
Variation of the frame yields the same result as variation of the metric. This is so because we have forced the torsion to vanish.

In dimension $D=2p$, ${\cal L}_{(p)}$ is the Euler density whose integral over the (compact) manifold is the Euler number~\cite{ChoDew82}~\cite{KobNom69}. In dimension $D>2p$,
${\cal L}_{(1)}, ..., {\cal L}_{(p)}$, sometimes called ``dimensionally continued" Euler densities, become ``dynamical" in the sense that their Euler variations, to wit the Lovelock
forms $E_{(1)A}, ..., E_{(p)A}$ are no longer identically zero.

For each term (27) we introduce the Sparling forms
$\sigma_{(p)A}$ and $\tau_{(p)A}$ defined as
$$\sigma_{(p)A} = - \omega^{BC} \theta^*_{ABCI_3 \cdots I_{2p}} (\Omega^{p-1})^{I_3 \cdots I_{2p}}\eqno({\rm II}.30)$$
$$\tau_{(p)A} = (\omega_A{}^B \omega^{CD}\theta^*_{BCDI_3 \cdots I_{2p}}- \omega^B{}_D \omega^{DC}\theta^*_{ABCI_3 \cdots I_{2p}})(\Omega^{p-1})^{I_3 \cdots
I_{2p}}.\eqno({\rm II}.31)$$
It can be shown that 
$$E_{(p)A} = \tau_{(p)A} - d\sigma_{(p)A}\,.\eqno({\rm II}.32)$$
 In form formalism this
result is due to Sparling. A proof can be found elsewhere~\cite{DubMad87}.
In the case $p=1$ the 3-form and the 2-form can be expressed in terms of
Christoffel symbols and become respectively the `pseudo-tensor' and the
`pseudo-potential'. Equation~(32) becomes the conservation of the
pseudo-tensor in vacuo.

Once one has written the Lovelock action in frame formalism it is an easy task to show  that the field equations are of second order but, apart in the $D\leq4$ case,   only {\it
quasi}-linear,  in that the  coefficients can contain derivatives.  We have shown in fact in (31) that each contribution to the Sparling $(D-1)$-form $\tau_A$ is the product
of a term quadratic in the connection and a power of the curvature. Equation~(32) states that the field equations are equivalent to the condition that this form be closed. Since the
exterior derivative of the curvature factors do not contribute, because of the Bianchi identities, the equations are of second order, except for the limiting value of $p$, in which
case they become identities. Because however of the curvature factors the chacteristic surfaces can be more complicated than a simple light
cone~\cite{Cho88a}~\cite{Ara88}\footnote{For example,  the propagation of high frequency metric perturbations  $h_{AB}=\varepsilon_{AB}e^{{\rm
i}\varphi}$ such that
$D_{AB}h_{CD}\approx \xi_A\xi_Bh_{CD}$ with
$\xi_A\equiv\partial_A \varphi$ and solution of $\delta \sigma^A_{(p)B}=0$ (N.B.~: {\it not} $\Sigma_p\alpha_p\lambda^{2(p-1)}\delta \sigma^A_{(p)B}=0$) is  determined by the
vanishing of the determinant~:
$\Delta=\xi^2L_{(p-1)}\sigma^{AB}_{(p-1)}\xi_A\xi_B$. (see ref (8) in ~\cite{Cho88a})}. 

\medskip

In the following we shall content ourselves with simple examples taken from cosmological models showing the kind of problems which may arise because of the quasi-linearity 
of the Lovelock Lagrangian and tensor.

\bigskip

\noindent
{\csc III. Examples from Kaluza-Klein Cosmology}

\medskip

Consider the situation when the $D$-dimensional manifold is the product of an ``external" ($d+1)$-dimensional Friedmann-Lema\^\i tre-Roberston-Walker spacetime and a $n$-dimensional
 compact ``internal" space of constant curvature. In an 
appropriate coordinate system the line element can be written as
$$ds^2=-dt^2+g_{\mu\nu}dx^\mu dx^\nu+g_{ab}dx^adx^b$$
$$\ \ \ =-dt^2+a^2_d(t)\,d\Omega^2_d+a^2_n(t)\,d\Omega_n^2\eqno({\rm III}.1)$$
where $d\Omega^2_{d,n}$ describe  ($d,n$)-dimensional maximally symmetric spaces. It is then a matter of straightforward 
calculation to obtain the components of the Lovelock tensor (II.19). We shall write them as
$$\sigma^0_{[D]0}=-{1\over2}F\qquad,\qquad\sigma^\mu_{[D]\nu}=-{1\over2}\delta^\mu_\nu\left[f_d+{\ddot a_d\over a_d}g_d+{\ddot a_n\over a_n}h_d\right]$$
$$ \sigma^a_{[D]b}=-{1\over2}\delta^a_b\left[f_n+{\ddot a_d\over a_d}h_n+{\ddot a_n\over a_n}g_n\right]\eqno({\rm III}.2)$$
where the coefficients $F$, $f_{d,n}$, $g_{d,n}$, $h_{d,n}$ are polynomials in  $h_{d,n}\equiv 
\lambda\dot a_{d,n}/ a_{d,n}$ and
$A_{d,n}\equiv\lambda^2k_{d,n}/a_{d,n}^2+h_{d,n}^2$ with $k_{d,n}$ taking the values ($+1, 0, -1$) and a dot representing $d/dt$. Their explicit
expressions have been obtained up to cubic order by M\"uller-Hoissen~\cite{Mul86}. They can be generalized to
any order when either both spaces are static or both spatially flat (see Deruelle and Fari$\tilde{\rm n}$a-Busto
 in~\cite{DerMad86a}). The field equations
$$\sigma^A_{[D]B}=T^A_B\eqno({\rm III}.3)$$
with $T^A_B$ the components of the stress-energy tensor of a perfect fluid, read
$$\rho={1\over2}F\quad,\quad {\ddot a_d\over a_d}g_d+{\ddot a_n\over a_n}h_d=-(f_d+\mu_d F)\quad,\quad
{\ddot a_d\over a_d}h_n+{\ddot a_n\over a_n}g_n=-(f_n+\mu_n F)\eqno({\rm III}.4)$$
where $T_{00}\equiv\rho$ is the energy density of matter and $\mu_{d,n}$ its adiabatic indices in the external/
internal spaces (defined by $T^\mu_\nu\equiv\mu_e\rho\delta^\mu_\nu$, $T^a_b\equiv\mu_e\rho\delta^a_b$).

These equations can serve as a basis to study various vacuum ($\rho=\mu_{d,n}=0$) states and their stability. For example, if
we impose the external space to be Minkowski spacetime ($d=3, k_d=0, x_d=0$), then the coefficients $\alpha_p$ must
satisfy two constraints (see e.g.~\cite{Mad85a}~\cite{Mul85}~\cite{DerMad86a})
$$\sum_{0\leq p< D/2}\alpha_p{n!\over(n-2p)!}\,{k_n^p\over a_n^{2p}}=0\quad,\quad \sum_{0\leq p< D/2}\alpha_p{(n-1)!\over(n-1-2p)!}\,{k_n^p\over a_n^{2p}}=0\,.\eqno({\rm
III}.5)$$ The first gives the radius of the internal space in terms of its dimension $n$, curvature $k_n$ and the parameters $\alpha_p$~; the second gives $\alpha_0$, say, in
terms of the other $\alpha_p$. Of course the solution may not be unique. The zero mode linear stability of
these ground states, that is, the behaviour of the solutions of (4) when expanded at linear order, was studied in~\cite{DerMad86a} and shown to impose further constraints on the
parameters
$\alpha_p$. (Linear stability of the other modes  was examined by Ishikawa~\cite{BouDes85}.)

The full-fledged equations (4) can also serve to build cosmological models. Indeed, when $d=3$
 and the size of the internal space is small enough, such a geometry is a potentially realistic candidate for describing
today's universe. However the (numerical) analysis of the dynamical evolution from an initial ground state where both the internal and external spaces
are of Planckian size to a final state where the external universe evolves in a quasi-Friedman way and the internal space
freezes out, has not been performed. An important word of caution is that if the sign of the determinant
$${\rm Det}\equiv g_ng_d-h_nh_d\eqno({\rm III}.6)$$
is positive, say, near the initial state and negative near the final state, then no dynamical evolution from one to the other is possible since the system (4) exhibits a breakdown
of predictability when the determinant goes through zero.

To illustrate this kind of possible pathology due to the quasi-linearity of the equations, consider as an example the case 
$d=4$ (with $k_4=0$), $n=0$. Then the field equations reduce to the quadrature
$$(\mu +1){t\over\lambda}=-6\int{\alpha_1+4\alpha_2\,h^2\over\alpha_0+12\alpha_1\,h^2+24\alpha_2\,h^4}\, dh\,.\eqno({\rm III}.7)$$
As studied by  Deruelle and Fari$\tilde {\rm n}$a-Busto~\cite{DerMad86a} universes obeying such an evolution law can exhibit
pathological behaviours such as ``coming into being" at a time $t=-t_1$ which is not a curvature singularity and then, the
solution of (7) being multivalued, having the possibility of either end up at a curvature singularity at some $t=-t_0$
or into ``nothingness" at $t=+t_1$ without any divergence in the curvature invariants signaling the approach of that 
event. The origin of
such a pathology lies in that the numerator of (7), that is, the coefficient of $\ddot a$ in the field equation which
determines the evolution of the scale factor, vanishes at $\pm t_1$. Hence, starting from some initial data, at $t=0$ for
example, the equation for $a(t)$ cannot predict its evolution beyond $t=t_1$. (See also, e.g.~\cite{Mul85}, ~\cite{Whe86} and, in the more recent context of
braneworlds, the ``Class I" solutions of~\cite{BinChaDav02}.)

These pathological cosmological models are just particularly simple examples of the non-invertibility of the operator 
$[\alpha_1(\gamma^{\mu\alpha}\gamma^{\nu\beta}-\gamma^{\mu\nu}\gamma^{\alpha\beta)}+\alpha_2 N^{\mu\alpha\nu\beta}]$, with 
$N^{\mu\alpha\nu\beta}$  defined in (II.13). We note that rewriting the second order terms in the Lanczos tensor under the form (II.14) is of not help to solve the ``Cauchy problem" 
since the first derivative terms hidden in the ($...)$ do not combine in an expression proportional to the term in $\{\}$.

\bigskip

\noindent
{\csc IV. Examples from ``brane worlds"}

\medskip 

A geometer can construct a ``braneworld" as follows~: consider a 5-dimensional spacetime $V_+$ with an edge (an easy way to vizualise this is to imagine, say,
the outside surface  of a half 2-sphere)~; make a copy $V_-$ of $V_+$ and superpose the copy and the original spacetimes onto each other along the edge (this is the
so-called
$Z_2$ symmetry)~; one thus obtains a spacetime, or ``bulk", $V_5$, without an edge (e.g. the outside {\it and} inside surfaces of a half 2-sphere) but which possesses a
singular surface, or ``brane"
$\Sigma_4$ whose extrinsic curvature is discontinuous~: the extrinsic curvature of $\Sigma_4$ embedded in $V_-$ (e.g. the interior of the 2-sphere) is the opposite of the
extrinsic curvature of $\Sigma_4$ embedded in $V_+$ (e.g. the exterior of the 2-sphere). 

Suppose now that the curvature of $V_+$ satisfies the vacuum Lanczos-Gauss-Bonnet equations everywhere, except on the edge, that is, is such that
$$\sigma^A_{[2]B}\equiv\Lambda \delta^A_B+\sigma^A_B+\alpha\, \sigma^A_{(2)B}=0\eqno({\rm IV}.1)$$
the Einstein and Lanczos tensors being defined in (II.7) (II.16).

Suppose also that $V_+$ is an anti-de Sitter spacetime everywhere but on the edge. Then, because of maximal symmetry,
$$R_{ABCD}=-{1\over {\cal L}^2}(g_{AC}g_{BD}-g_{AD}g_{BC})\eqno({\rm IV}.2)$$
with
$${1\over {\cal L}^2}={1\over4\alpha}\left(1\pm\sqrt{1+{4\alpha\Lambda\over3}}\right)
\eqno({\rm IV}.3)$$
because of (1). (One usually chooses the $-$  sign so that, when $\alpha\to0$, ${\cal L}^2\to-{6\over\Lambda}$, that is the Einsteinian value.)

Finally, suppose, for the time being and the sake of the example, that $\Sigma_4$ is flat. To find
the appropriate edge of $V_+$ (which is locally AdS), one describes it in quasi-conformal coordinates such that its line element reads
$$ds^2=dw^2+e^{-2w/{\cal L}}\eta_{\mu\nu}dx^\mu dx^\nu\eqno({\rm IV}.4)$$
(${\cal L}>0$) and one keeps only the $w>0$ part, in order that the edge $w=0$ be flat. One then describes $V_-$ with  $w<0$, so that its line element reads
$ds^2=dw^2+e^{+2w/{\cal L}}\eta_{\mu\nu}dx^\mu dx^\nu$.

Consider now the complete space-time $V_5\ {\cal t}\ \Sigma_4$. It is locally anti-de Sitter and solution of (1) everywhere but on $\Sigma_4$. In the chosen coordinates its line
element is
$$ds^2=dw^2+e^{-2|w|/{\cal L}}\eta_{\mu\nu}dx^\mu dx^\nu\,.\eqno({\rm IV}.5)$$
The extrinsic curvature of $\Sigma_4$ is discontinuous (it is $K_{\mu\nu}={1\over{\cal L}}\eta_{\mu\nu}$ on the $V_+$ side and $K_{\mu\nu}=-{1\over{\cal L}}
\eta_{\mu\nu}$ on the $V_-$ side), that is
$${\cal K}^\mu_\nu={1\over{\cal L}}\delta^\mu_\nu\,{\cal S}(w)\eqno({\rm IV}.6)$$
where the sign distribution  ${\cal S}(w)$ is +1 if $w>0$, and -1 if $w<0$. Some components of the Riemann tensor therefore exhibit a delta-type singularity (since
${\cal S}'(w)=2\delta(w)$) and one expects that $V_5\ {\cal t}\ \Sigma_4$ satisfies the Lanczos-Gauss-Bonnet equations everywhere---that is, $\Sigma_4$ included--- but in
the presence of ``matter" localised on
$\Sigma_4$, i.e. that one has
$$\sigma^A_{[2]B}= T^A_B\,{\cal D}(w)\eqno({\rm IV}.7)$$
where $T^A_B$ is interpreted as the ``stress-energy" tensor of matter in the brane and where ${\cal D}$ is a distribution localized on $\Sigma_4$, i.e. proportional to some linear
 combination of the Dirac delta distribution and its derivatives (see
e.g.~\cite{ChoDew82}).

\medskip

The question now is to express this ``stress-energy" tensor in terms of the discontinuity of the extrinsic curvature.

Since the line element (5) is written in Gaussian-normal coordinates, the ($4+1$) decomposition (II.12) of the components of the Riemann tensor, together with the expression (6)
for the extrinsic curvature give, near $\Sigma_4$
$$R_{w\mu w\nu}=\left[{2\delta(w)\over{\cal L}}+{{\cal S}^2(w)\over{\cal L}^2}\right]
\eta_{\mu\nu}\ ,\  R_{w\mu\nu\rho}=0\ ,\ 
R_{\mu\nu\rho\sigma}=\left[{{\cal S}^2(w)\over{\cal L}^2}\right](\eta_{\mu\rho}\eta_{\nu\sigma}-
\eta_{\mu\sigma}\eta_{\nu\rho}).\eqno({\rm IV}.8)$$
Note that ${\cal S}^2(w)=1$ everywhere but on $w=0$ and is not straightforwardly defined in a distributional sense (it is however not
``dangerous" when multiplied by just a constant). As for the term proportional to
$\delta(w)$, it is distributionally very  defined. The Einstein tensor  reads
$$\sigma^w_w=6{{\cal S}^2(w)\over{\cal L}^2}\quad,\quad \sigma^w_\mu=0\quad,\quad \sigma^\mu_\nu=-6\,\delta^\mu_\nu
\left[{\delta(w)\over{\cal L}}-{{\cal S}^2(w)\over{\cal L}^2}\right]\,.\eqno({\rm IV}.9)$$

Hence, in pure Einstein theory ($\alpha=0$, ${6\over{\cal L}^2}=-\Lambda$) one recovers the well-known result~\cite{RanSun99a}
$$T^w_w=T^w_\mu=0\quad,\quad T^\mu_\nu=-{6\over{\cal L}}\delta^\mu_\nu \eqno({\rm IV}.10)$$
which is nothing but the ``junction conditions"~\cite{Isr66} applied to the problem at hand.

Now, the Lanczos tensor $\sigma^A_{(2)B}$ is only quasi-linear in the second derivatives of the metric that is contains, when calculated in gaussian-normal coordinates, terms of
the  $K\,.\,K\,.\,{\partial K\over\partial w}$ type, see (II.13). More precisely, for an anti-de Sitter bulk and flat brane, (II.13) gives
$$\sigma^\mu_{(2)\nu}={24\over{\cal L}^3}\,\delta^\mu_\nu\,{\cal S}^2(w)\delta(w)+...\eqno({\rm IV}.11)$$
where the dots symbolize non dangerous terms.

Since the product of the Dirac and sign distributions is not straightforwardly defined, various proposals have been put forward to give a meaning to (11) [38-55].

Some, starting from (11), have defined ${\cal S}^2=1$ everywhere, including $w=0$ (despite the fact that, then,  ${\cal S}'=0$ and not $2\delta$) and hence rewritten it as
$$\sigma^\mu_{(2)\nu}\simeq{24\over{\cal L}^3}\,\delta^\mu_\nu\,\delta(w)+...\eqno({\rm IV}.12)$$
This way, they obtained that the stress-energy tensor of the brane is
$$ T^\mu_\nu\simeq -{6\over{\cal L}}\,\delta^\mu_\nu\,\left(1-{4\alpha\over{\cal L}^2}\right)\,.\eqno({\rm IV}.13)$$

Others started from  ${\cal S}'=2\delta$ and the Leibniz rule ${\cal S}^2{\cal S}'\simeq{1\over3}({\cal S}^3)'$ (which is what the rewritting of (II.13) as (II.14) reduces to in this
case\footnote {However, according to some authors, the Leibniz rule may not apply to products of distributions.}). They then followed the same procedure as
before to write
${1\over3}({\cal S}^3)'={1\over3}({\cal S}^2{\cal S})'={1\over3}{\cal S}'$
 so that (11) becomes\footnote{Indeed, if
$\phi(w)$ is a test function, then $\int_{-\infty}^{+\infty}dw\,\phi(w)(S^3)'=2\phi(0)$ and hence $(S^3)'=2\delta(=S')$.}
$$\sigma^\mu_{(2)\nu}\simeq{8\over{\cal L}^3}\delta^\mu_\nu\,\delta(w)+...\eqno({\rm IV}.14)$$
and hence obtained the following brane stress-energy tensor
$$T^\mu_\nu\simeq -{6\over{\cal L}}\,\delta^\mu_\nu\, \left(1-{4\alpha\over3{\cal L}^2}\right)\,.\eqno({\rm IV}.15)$$

Finally Deruelle and Dolezel~\cite{DerDoz00} claimed that the fact that (11) is ill-defined signals that one  may not treat $\Sigma_4$ as a ``thin shell" and that the ``junction
conditions" may depend on the microphysics, that is the way one represents it as a ``thick" shell. Then the stress energy tensor of a flat brane was advocated to be
$$T^\mu_\nu= -{6\over{\cal L}}\delta^\mu_\nu\left(1-A{4\alpha\over{\cal L}^2}\right)\eqno({\rm IV}.16)$$
 $A$  being a constant which encapsulates the microphysics of the brane.

When the brane is flat, the difference between the various approaches is immaterial as it amounts to a renormalisation of the brane tension. But it matters when one treats
cosmological models.

To find a Friedmann-Lema\^\i tre- Robertson-Walker surface $\Sigma_4$ in a five dimensional anti-de Sitter spacetime, 
one can introduce various coordinate systems adapted to the problem and write the line element (1) under, e.g., its static form
$$ds^2=-\left(k+r^2/{\cal L}^2\right)dt^2+{dr^2\over k+r^2/{\cal L}^2}+r^2\,d\Omega^2_{(3)}\eqno({\rm IV}.17)$$
with $k=0$ or $\pm1$ characterizing the curvature of the maximally symmetric 3-space of line element $d\Omega^2_{(3)}=
d\chi^2+f^2_k(d\theta^2+\sin^2\theta\,d\varphi^2)$ ($f_0=\chi$, $f_1=\sin\chi$, $f_{-1}=\sinh\chi$). The FLRW surface  $\Sigma_4$ is then defined by
$$r=a(\tau)\quad,\quad t=t(\tau)\quad\hbox{with}\quad \dot t=a{\sqrt{h^2+k/a^2+1/{\cal L}^2} \over k/a^2+1/{\cal L}^2}\eqno({\rm IV}.18)$$
$a(\tau)$ being the scale factor and $h=\dot a/a$. In the Gaussian normal coordinate system first introduced by Binetruy et al.~\cite{BraBru03}
$$ds^2=dy^2-n^2(\tau,y)d\tau^2+S^2(\tau,y)\,d\Omega^2_{(3)\,,}\eqno({\rm IV}.19)$$
where the expressions for $n(\tau,y)$ and $S(\tau,y)$ can be found in~\cite{BraBru03}, the equation for $\Sigma_4$ is simply $y=0$.

The relevant component of the extrinsic curvature of $\Sigma_4$ in $V_+$ is computed to be
$${\cal K}\equiv K^\chi_\chi=K^\theta_\theta=K^\varphi_\varphi=-{S'\over S}=\sqrt{h^2+{k\over a^2}+{1\over{\cal L}^2}}\eqno({\rm IV}.20)$$
a prime representing ${\partial\over\partial y}$. Introducing the distribution
$${1\over{\cal L}^2(y)}\equiv{1\over {\cal L}^2}+ {\cal K}^2\left[{\cal S}^2(y)-1\right]\eqno({\rm IV}.21)$$
(which is equal to the constant $1/{\cal L}^2$ everywhere but at $y=0$) it is straightforward to obtain, using either (II.13) or (II.14), the Einstein-Lanczos tensor,
which is proportional to the Dirac distribution~:
$$\sigma^\tau_{[2]\tau}=-6{\cal K}\delta(y)\left[1-{4\alpha\over{\cal L}^2(y)}\right]\eqno({\rm IV}.22)$$
or
$$\sigma^\tau_{[2]\tau}=-3{\cal K}\left\{{\cal S}(y)\left[1-{4\alpha\over{\cal L}^2(y)}+{8\alpha\over3}{\cal K}^2
{\cal S}^2(y)\right]\right\}'\,.\eqno({\rm IV}.23)$$
Just as in the case of a flat brane seen above, neither expression (22) nor (23) is straightforwadly  defined in a distributional sense. If now, as before, one defines ${\cal S}^2=1$
everywhere including
$\Sigma$ in either (22) or (23) one gets the following, very different, results
$$\rho\simeq 6\left(1-{4\alpha\over{\cal L}^2}\right)\sqrt{h^2+{k\over a^2}+{1\over{\cal L}^2}}\eqno({\rm IV}.24)$$
or
$$\rho\simeq6\sqrt{h^2+{k\over a^2}+{1\over{\cal L}^2}}\left[1-{4\alpha\over{\cal L}^2}+{8\alpha\over3}
\left(h^2+{k\over a^2}+{1\over{\cal L}^2}\right)\right]\,.\eqno({\rm IV}.25)$$
The first result (24), based on (II.13) and advocated by e.g~\cite{GerSop02}, implies that modifying Einstein's 
gravity by the inclusion of the Gauss-Bonnet correction amounts to nothing more (in this particular case) than a coupling-renormalisation. On the other hand, the second result (25)
 based on (II.14) and  advocated by, e.g.~\cite{BinChaDav02}
yields a very different cosmological evolution (see, e.g. Lidsey and Nunez in~\cite{LidNojOdi02}).

In the general case, the issue is then whether or not the stress-energy tensor of matter in the brane, as deduced from
 (II.14), that is
$$T^\mu_\nu=2\left[K^\mu_\nu-K\delta^\mu_\nu+4\alpha\left({3\over2}J^\mu_\nu-{1\over2}J\delta^\mu_\nu-
\bar P^\mu_{\ \rho\nu\sigma}K^{\rho\sigma}\right)\right]\eqno({\rm IV}.26)$$
($K^\mu_\nu$ being the extrinsic curvature of the brane in $V_+$), which reduces to (25), is correct or not.

\medskip

Recently Davis~\cite{Dav02} and Gravanis and Willison~\cite{GraWil02}, using a former result by Myers~\cite{Mye87} (see also~\cite{GrePad03}),
 proposed to settle the issue positively by considering more closely the boundary terms which arise when varying the 
Lovelock  Lagrangian with respect to the metric. As we have seen in Section II the variation of the Hilbert-Lanczos 
action is
$$\delta S\equiv \delta\int_{{\cal M}}\!d^5x\,\sqrt{-g}\,L_{[2]}=\int_{{\cal M}}\!d^5x\,\sqrt{-g}\,\sigma^{[2]}_{AB}\,\delta g^{AB}
+\int_{\partial{\cal M}}\!d^4x\,\sqrt{-\gamma}\,n_C\, (V^C_{(1)}+\alpha V^C_{(2)})\eqno({\rm IV}.27)$$
with $V^C_{(2)}$ given, up to a divergenceless vector by, e.g. (II.9), and $V^C_{(1)}$  given in footnote 1~; $\gamma_{\mu\nu}$ is the induced metric on $\partial{\cal M}$ and $n^C$ the
unit vector pointing outside the boundary. If, now, one imposes that the field equations be obtained by keeping the metric coefficients alone fixed at the boundary (their
derivatives remaining free), an extra term must be added to the Lagrangian.  Now, the boundary term can be written as
$$\sqrt{-\gamma}\,n_C\,(V^C_{(1)}+\alpha V^C_{(2)})=\delta(\sqrt{-\gamma}\,Q)+\sqrt{-\gamma}\,B_{\mu\nu}\delta\gamma^{\mu\nu}+\alpha\sqrt{-\gamma}\nabla_\rho
W^\rho\eqno({\rm IV}.28)$$
where only the vector $W^\rho=W^\rho(\delta\gamma_{\mu\nu}, \nabla_\sigma\delta\gamma_{\mu\nu})$ depends upon the choice of the divergenceless vector defining $V^C_{(2)}$.
 The variation of the action can then be cast into the form (keeping $\delta\gamma_{\mu\nu}$ and $\nabla_\rho\delta\gamma_{\mu\nu}$ fixed at the boundaries of $\partial M$)
$$\delta S_b\equiv\delta\left(\int_{{\cal M}}\!d^5x\sqrt{-g}\,L_{[2]}-\int_{\partial{\cal M}}\!d^4x
\sqrt{-\gamma}\,Q\right)=$$
$$=\int_{{\cal M}}\!d^5x\sqrt{-g}\,\sigma^{[2]}_{AB}\,\delta g^{AB}+\int_{\partial{\cal M}}\!d^4x\sqrt{-\gamma}\,B_{\mu\nu}\,\delta\gamma^{\mu\nu}.\eqno({\rm IV}.29)$$
The Chern-Simons form $Q$ and $B_{\mu\nu}$ can be written in the language of forms, and their explicit expressions are~\cite{Mye87},~\cite{Dav02},~\cite{GraWil02}
$$Q=2K+4\alpha(J-2\bar\sigma_{\mu\nu}K^{\mu\nu})\eqno({\rm IV}.30)$$
$$B_{\mu\nu}=K_{\mu\nu}-K\gamma_{\mu\nu}+4\alpha\left({3\over2}J_{\mu\nu}-{1\over2}J\gamma_{\mu\nu}-
\bar P_{\mu\rho\nu\sigma}K^{\rho\sigma}\right).\eqno({\rm IV}.31)$$
Since $B_{\mu\nu}$ is nothing but (26), one can argue
convincingly that indeed, (26) is the correct junction condition.
More precisely, taking the total action as $S=S_b+\int_{\Sigma_4}d^4x\sqrt{-\gamma}L_m$ with $L_m(\gamma_{\mu\nu})$ the brane matter Lagrangian whose Euler variation
yields the brane stress-energy tensor $\delta(\sqrt{-\gamma}L_m)\equiv -{1\over2}\sqrt{-\gamma}T_{\mu\nu}\delta\gamma^{\mu\nu}$, we get (26) as the brane gravity
equations.

\bigskip

{\bf Acknowledgements~:} 
One of us (N.D.) thanks Joseph Katz and Cristiano Germani for fruitful  discussions about matching conditions in gravity theories, and the Yukawa Institute for Theoretical
Physics where this work was completed,  for its hospitality.

\providecommand{\href}[2]{#2}\begingroup\raggedright\endgroup

\end{document}